\documentclass[%
preprint,
superscriptaddress,
amsmath,amssymb,
aps,
longbibliography,
]{revtex4-1}
\usepackage{amsmath}
\usepackage{amssymb}
\usepackage{graphicx}
\usepackage{braket}
\usepackage{color}
\usepackage{xcolor}
\usepackage{siunitx}
\usepackage{upgreek}
\usepackage[utf8]{inputenc}
\usepackage{physics}
\usepackage{multirow}
\usepackage{array}
\usepackage{url}
\usepackage{dcolumn}
\usepackage{bm}
\usepackage{hyperref}
\usepackage{natbib}
\usepackage{setspace}

\begin{document}



\title{Hyperfine-mediated transitions between electronic spin-1/2 levels of transition metal defects in SiC}

\author{Carmem~M.~Gilardoni}
\email{c.maia.gilardoni@rug.nl}
\affiliation{Zernike Institute for Advanced Materials, University of Groningen, NL-9747AG  Groningen, The Netherlands}
\author{Irina~Ion}
\affiliation{Zernike Institute for Advanced Materials, University of Groningen, NL-9747AG  Groningen, The Netherlands}
\author{Freddie~Hendriks}
\affiliation{Zernike Institute for Advanced Materials, University of Groningen, NL-9747AG  Groningen, The Netherlands}
\author{Michael~Trupke}
\affiliation{Vienna Center for Quantum Science and Technology, University of Vienna, VCQ, Boltzmanngasse 5,1090 Vienna, Austria}
\author{Caspar~H.~van~der~Wal}
\email{c.h.van.der.wal@rug.nl}
\affiliation{Zernike Institute for Advanced Materials, University of Groningen, NL-9747AG  Groningen, The Netherlands}

\date{Version of \today}


\begin{abstract}
Transition metal defects in SiC give rise to localized electronic states that can be optically addressed in the telecom range in an industrially mature semiconductor platform.
This has led to intense scrutiny of the spin and optical properties of these defect centers.
For spin-1/2 defects, a combination of the defect symmetry and the strong spin-orbit coupling may restrict the allowed spin transitions, giving rise to defect spins that are long lived, but hard to address via microwave spin manipulation.
Here, we show via analytical and numerical results that the presence of a central nuclear spin can lead to a non-trivial mixing of electronic spin states, while preserving the defect symmetry.
The interplay between a small applied magnetic field and hyperfine coupling opens up magnetic microwave transitions that are forbidden in the absence of hyperfine coupling, enabling efficient manipulation of the electronic spin.
We also find that an electric microwave field parallel to the c-axis can be used to manipulate the electronic spin via modulation of the relative strength of the dipolar hyperfine term.
\end{abstract}

\maketitle
\setstretch{1}
\section{Introduction}

Solid state defects and molecular centers with optically addressable spin provide versatile platforms for the implementation of quantum technologies \cite{Awschalom2013, atature2018, Gaita-Arino2019, Coronado2020,Wolfowicz2021_Rev,Chatterjee2021,Awschalom2021,Castelletto2020}.
When placed in a crystal lattice, transition metal (TM) defects with active spins show strong analogy to molecular metal-organic centers \cite{Atzori2016,Wojnar2020,Gimenez2020,Diler2020,bosma2018,gallstrom2012,maier1992,baur1997,Yamabayashi2018}.
Their spin properties arise from electrons strongly localized at the $d$-orbitals of the TM core, subject to the relevant point-group symmetries.
The V and Mo defects in hexagonal SiC have been extensively investigated recently, motivated by the fact that their optical transitions are near or at telecom-standard wavelengths, and that SiC is an industrially mature semiconductor platform \cite{bosma2018,Gilardoni2020,Csore2020,Tissot2021Spin,spindlberger2019,Wolfowicz2020}.
When these defect centers are in a charge state with one localized $d$-electron, they are largely analogous to each other, both having ground states with electronic spin-$1/2$ behavior.

Group theory predicts limited possibilities for driving spin transitions in these defect centers at moderate magnetic fields, compatible with the long spin lifetimes observed for Mo defects \cite{Gilardoni2020,Tissot2021Spin,Csore2020}.
Nevertheless, experimental results \cite{baur1997,Wolfowicz2020,kaufmann1997,maier1992} demonstrate efficient electron spin resonance on V defects with an oscillating magnetic field parallel to the symmetry axis.
These results seem inconsistent, and have puzzled the community.
Phenomenological interpretations of experiments \cite{Wolfowicz2020} have considered a role for the central nuclear spin, but these are incompatible with the spatial symmetry of the defect.

Here, we develop a bottom-up analytical and numerical model of a single electron in the $d$-orbitals of the TM core, and include hyperfine interaction with the central nuclear spin.
We focus on the energy levels and transitions between ground-state spin-$1/2$ levels of these defects, and find strong hyperfine-mediated matrix elements for transitions that are forbidden in a description without hyperfine coupling.
This sheds light on the experimentally observed magnetic microwave driving.
The anisotropic dipolar interaction between magnetic moments of the electron and nucleus leads to mixing between the electron spin sublevels, which enables efficient electron paramagnetic resonance with oscillating magnetic fields parallel to the symmetry axis of the defect centers.
In addition, we find that the mechanism of hyperfine-mediated spin resonance also allows for efficient driving with the electric-field component of microwave fields.
An oscillating electric field along the growth axis of the crystal -- a technologically relevant geometry with multiple fabrication possibilities \cite{Klimov2014,falk2014,Anderson2019,Widmann2019,Lukin2019} -- also leads to hyperfine-mediated electron spin transitions.
While preserving the original uniaxial symmetry of the defects, these electric fields modulate the spatial distribution of electron density in the ground state, which in turn strongly influences the anisotropic dipolar interaction between nuclear and electronic spins.
For specific relevance to V and Mo in SiC, we provide effective-spin Hamiltonians for the ground-state doublets in terms of parameters relating directly to the configuration of the defect centers.
In this way, our approach provides an easy way to interpret these fitting parameters in terms of the microscopic configuration of the defect centers, and is complementary to purely group-theoretical methods \cite{Tissot2021Hyperf} that yield symmetry-restricted effective-spin Hamiltonians that can be fit to experiments.
Additionally, our approach is also complementary to results from \textit{ab initio} calculations of the defects' geometric and electronic structure \cite{Csore2020}: here, we include the different parts of the Hamiltonian one by one, and this allows us to determine what is the particular interplay of different interactions that leads to the experimentally observed behavior.

\subsection{Review of literature and defect configuration}
\label{Sec::Intro}

The Mo and V defects in hexagonal SiC that are optically active in the near infrared range are composed of a TM impurity substituting a Si atom in the lattice.
Spectroscopic studies show that both ground and optically excited state manifolds consist of two and three, respectively, sets of Kramers doublets (KD, effective spin-$1/2$ doublets), as in the right-panel schematic of figure~\ref{Fig::Model}(b) \cite{bosma2018,maier1992,kaufmann1997,Wolfowicz2020,spindlberger2019}.
The two KDs in the ground-state manifold show strongly anisotropic interaction with magnetic fields.
In particular, for the spin doublet with lowest energy, the spin-1/2 degeneracy is easily broken by a magnetic field parallel to the crystal c-axis (which coincides with the three-fold rotational axis of the defects), but a magnetic field perpendicular to this direction gives rise to a Zeeman splitting that is zero (or close to zero, depending on the particular lattice site occupied by the impurity).
These spectral features can in good approximation be explained as a defect center in a Si-substitutional site that, after bonding to the lattice (and ionization in the case of Mo), is left with one unpaired electron spin in the $d$-orbitals of the TM core.
In this case, the strongly anisotropic Zeeman spectrum is a consequence of a combination of the defect three-fold rotational symmetry and spin-orbit coupling \cite{bosma2018,kaufmann1997,Csore2020,Gilardoni2020,Tissot2021Spin}, which gives rise to two ground-state KDs transforming as the $\Gamma_4$ and $\Gamma_{5,6}$ irreducible representations of the double group $\bar{C}_{3v}$, and separated in energy due to spin-orbit coupling. These two doublets are spin-orbit split, such that they are respectively formed by states having orbital angular momentum antiparallel or parallel to the electron spin. As far as symmetry is concerned, a doublet transforming as $\Gamma_4$ transforms as states with spin projection along the symmetry axis $m_s = \pm1/2$. Thus, these states may have contributions from orbitals with zero angular momentum (see also \cite[Sec.~I]{SI}). In contrast, a doublet transforming as $\Gamma_{5,6}$ transforms as states with $m_s = \pm 3/2$. Thus, all orbitals that make up $\Gamma_{5,6}$ states have orbital angular momentum non-zero.

For KDs transforming as $\Gamma_{5,6}$, interaction with a magnetic field perpendicular to the rotational axis of the defect is strictly symmetry forbidden \cite{Gilardoni2020,Tissot2021Spin}.
Earlier work found it thus plausible that the KD of lowest energy should be the KD that transforms as $\Gamma_{5,6}$, due to the observed $g_\bot$ parameter being $0$ \cite{bosma2018,Gilardoni2020,maier1992,kaufmann1997}.
Progressing insight from this and parallel work \cite{Tissot2021Hyperf, Csore2020}, however, shows that the KD with lowest energy is in fact the $\Gamma_4$ KD after all.
Here, $g_\bot$ can also be near zero, but this only occurs for a certain parameter range for spin-orbit coupling and crystal-field strengths \cite{Tissot2021Spin}.
Spin doublets transforming as $\Gamma_4$ may thus also interact very weakly with magnetic fields perpendicular to the crystal c-axis, giving rise to near-zero Zeeman splittings, smaller than the accuracy of most experimental setups at magnetic fields below 1~T.
Thus, the $g$-parameter alone cannot be used to distinguish between KDs of the two kinds in this magnetic field range.
However, the two types of KDs are distinguishable via their hyperfine structure, since the effective-spin Hamiltonians associated with them differ significantly and give qualitatively different spectra, as recently observed \cite{Wolfowicz2020} (see below for more details).
This conclusion was obtained in parallel also using a purely group-theory approach \cite{Tissot2021Hyperf}.
Both V and Mo have naturally abundant isotopes with non-zero nuclear spin.
Whereas $25\%$ of stable molybdenum nuclei have spin-$5/2$ (the remaining isotopes have nuclear spin zero), over $99\%$ of naturally occurring vanadium nuclei have spin-$7/2$.
In this way, measurements on V ensembles at low magnetic fields \cite{Wolfowicz2020} provide direct access to the defects' hyperfine structure, which allows us to determine that the lowest energy KD of V defects in SiC transforms as $\Gamma_4$.

\section{Methods}

We model the spin properties of the system by considering a single electron in the $d$-orbitals of the TM core.
The $d$-orbitals of the TM core on itself are isolated in energy from other orbital states due to the Coulomb interaction with the nucleus, and are five-fold degenerate in the presence of spherical symmetry.
Via a Hamiltonian of the type

\begin{equation}
	H = H_{\text{crystal}} + H_{\text{SOC}} + H_{\text{HF}} + H_{\text{Zee}}
	\label{eq::H_all}
\end{equation}

\noindent we consider the crystal-field potential arising from interaction with the lattice, spin-orbit coupling, hyperfine interaction with the central nuclear spin and Zeeman interaction with an external magnetic field, respectively, acting on these states.

We note that the full description of the electronic wavefunctions must be obtained from first-principles calculations that take hybridization with nearby atomic orbitals and the core orbitals of the TM core into account \cite{Csore2020}.
However, modeling the ground-state KDs as an electron localized in the $d$-orbitals of the TM
provides a good approximation \cite{spindlberger2019,Csore2020,kaufmann1997,bosma2018,Gilardoni2020}. The approach of our work thereby encompasses and provides intuition into the relevant physical processes at play at low computational cost. Additionally, since we include the contributions to the Hamiltonian one by one, our approach also provides insight into how the combination of different interactions (crystal field, SOC, Zeeman and hyperfine) leads to non-trivial eigenstates.
When considering the optically excited states, the contribution from bonding with lattice states is more significant \cite{Csore2020,spindlberger2019}, and the validity of our model for these KDs is limited.
For this reason, we focus our analysis on the ground-state properties of these defect centers.

{\begin{figure*}[t]
		\includegraphics[width=\textwidth]{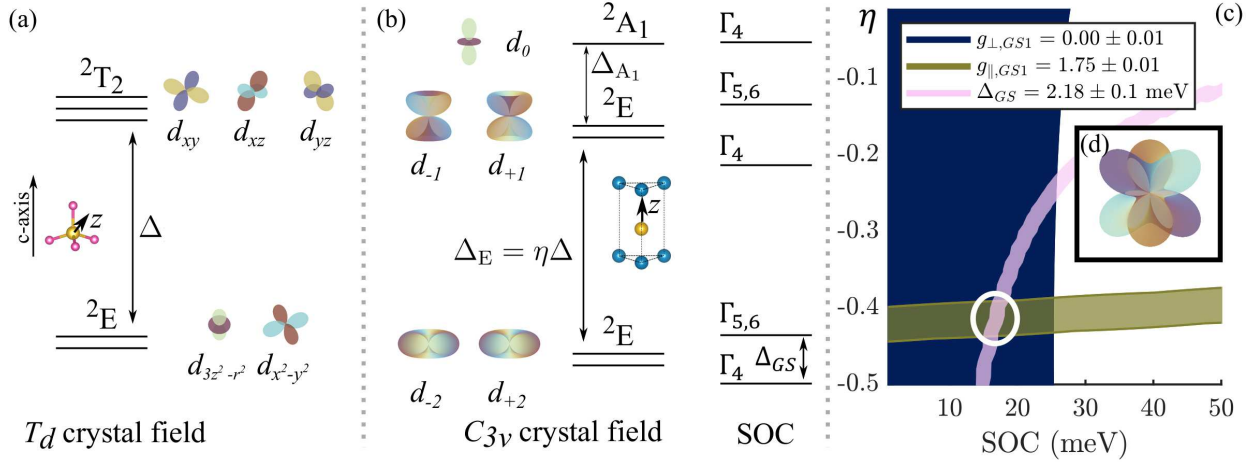}
		\caption{(a) Transition metal (TM) defects in hexagonal SiC in a substitutional site are composed of a central TM impurity (yellow in the inset) surrounded by nearest neighbors lattice atoms (pink in the inset) distributed in a tetrahedral configuration. This symmetry splits the five-fold degenerate $d$-orbitals of the transition metal core into a ground state orbital doublet and an excited state orbital triplet, separated by an energy $\Delta$. These orbitals are most intuitively expressed in terms of cubic harmonic functions, where the $z$-axis points between two of the tetrahedral bonds. (b) The second-nearest neighbors and third-nearest neighbors are distributed in a trigonal prismatic configuration, which splits the excited state triplet into a doublet and a singlet, and rotates the natural basis associated with the crystal field. The trigonal-prismatic contribution to the crystal field is most intuitively expressed in the basis of spherical harmonics with a $z$-axis along the three-fold rotational axis. This three-fold rotational axis coincides with the crystal c-axis. Spin-orbit interaction further splits both doublets into two Kramers doublets, each transforming as the irreps $\Gamma_4$ and $\Gamma_{5,6}$ of the double group $\bar{C}_{3v}$. (c) We model the TM defect in SiC by considering the contributions of both a tetrahedral crystal field and a trigonal crystal field, with relative strengths provided by the parameter $\eta$. By varying both $\eta$ and the strength of spin-orbit coupling, we can extract ranges where the model agrees with the experimentally measured energy splittings in the ground state (colored areas in the plot). (d) At the region where the model provides reliable spin-orbit and Zeeman energy splittings for the ground state KD (where all three shaded areas overlap, circled), the electronic wavefunctions of both ground state Kramers doublets have the electron density concentrated slightly above or below the transition metal ion. In the graphical representation of the wavefunctions, the color corresponds to a phase according to the scale bar in figure~\ref{Fig::Matrix}(b).
			\label{Fig::Model}}
	\end{figure*}
}

\subsection{Crystal field potential}
\label{Sec::Crystal}

A TM impurity in a Si-substitutional site is surrounded by carbon atoms arranged in a tetrahedral configuration around it, such that the strongest contribution to the crystal field has tetrahedral symmetry ($T_d$).
This crystal potential splits the five-fold degenerate $d$-orbitals into a ground state orbital doublet ($^2\text{E}$) and an excited state orbital triplet ($^2\text{T}_2$), separated by an energy splitting $\Delta$, and with natural quantization along the high symmetry axis $\hat{z}$ pointing between two of the tetrahedral bonds (figure~\ref{Fig::Model}(a)).
In hexagonal polytypes of SiC, the second and third nearest neighbors of the TM impurity are arranged in a trigonal configuration.
On their own, these lattice atoms give rise to a crystal field with $C_{3v}$ symmetry that splits the $d$-orbitals into 2 orbital doublets ($^2\text{E}$) separated by an energy $\Delta_{\text{E}}$ and an orbital singlet ($^2\text{A}_1$) separated from the top doublet by an energy $\Delta_{\text{A}_1}$.
In this case, the eigenstates are given by linear combinations of the spherical harmonics with $l = 2$, with $\hat{z}$ pointing along the three-fold rotational axis (see figure~\ref{Fig::Model}(b)).
Although the trigonal crystal field arises from the second and third nearest neighbors of the TM impurity, experimental results show that it is of significant importance for determining the spectral properties of the defect spin \cite{kaufmann1997,Csore2020}.
For this reason, we do not include it as a perturbation, but rather diagonalize the full crystal-field Hamiltonian with both trigonal and tetrahedral contributions.
We parameterize the energy term $\Delta_\text{E}$ arising from the trigonal crystal field in terms of the energy term $\Delta$ arising from the tetrahedral crystal field, such that $\Delta_\text{E} = \eta \Delta$ (figure~\ref{Fig::Model}), where $\eta$ provides the relative strength between the trigonal and tetrahedral crystal field contributions.
We can include the effect of electric fields or strain parallel to the three-fold rotational symmetry by considering deformations that preserve trigonal symmetry via small changes of $\eta$.

\subsection{Spin-orbit coupling}

We include the effect of spin-orbit coupling (in the same manner as Tissot \textit{et al.}, \cite{Tissot2021Spin}) via a Hamiltonian of the type

\begin{equation}
	H_{\text{SOC}} = \lambda k\mathbf{L}\cdot \mathbf{S}
\end{equation}

\noindent where $\lambda$ is the reduced matrix element indicating the strength of the spin-orbit interaction, $\mathbf{L}$ is the orbital angular momentum operator, which acts on the spatial part of the wavefunction, and $\mathbf{S}$ is the spin operator acting directly on the electronic spin. The parameter $k$ accounts for an empirical reduction of the orbital angular momentum of the electronic states due to bonding to nearby atoms or dynamic Jahn-Teller effects, for example.

In first-order, this spin-orbit coupling Hamiltonian gives rise to an additional splitting between the orbital doublets defined by the crystal field.
Within the ground state space, two Kramers doublets arise, transforming as $\Gamma_4$ and $\Gamma_{5,6}$ respectively, and separated by an energy $\Delta_{GS}$ (see figure~\ref{Fig::Model}(b)) \cite{Tissot2021Spin,bosma2018,maier1992,Gilardoni2020}.
Such a KD is an \textit{effective} spin-$1/2$ system.
Its overall spin properties depend on both the intrinsic spin of the electron and its orbital angular momentum \cite{Csore2020}.

\subsection{Hyperfine interaction}

We include the interaction with a central nuclear spin via a Hamiltonian of the type

\begin{equation}
	H_{\text{HF}} = A \Big(k\mathbf{L}\cdot \mathbf{I} + 3 \frac{(\mathbf{S}\cdot\vec{r})(\mathbf{I}\cdot\vec{r})}{r^2} - \mathbf{S}\cdot\mathbf{I}\Big)
\end{equation}

\noindent where $\vec{r}$ is the vector from nucleus to electron, and $\mathbf{I}$ is the nuclear spin operator. We define the term $A \equiv \frac{\mu_0}{4\pi} \frac{g_e \mu_B g_n \mu_N}{h^2}\frac{1}{r^3}$, where $\mu_0$ is the vacuum permeability, $g_e$ ($g_n$) is the electronic (nuclear) g-parameter, $\mu_B$ ($\mu_N$) is the electron (nuclear) magnetic moment and $h$ is Planck's constant.
This Hamiltonian includes the classical dipolar interaction between the magnetic moments of the nucleus and electron, as well as the effective magnetic field at the nucleus due to the orbital angular momentum of the electron. We do not include the electric quadrupolar interaction since, to first order, it does not lead to mixing between two states belonging to a KD \cite{abragam1970}.

Here, we do not take into account the Fermi contact interaction between nuclear and electron spins.
In an approach focused solely on the $d$-orbitals of the TM core, this term is zero due to the vanishing electron density at the nucleus associated with these orbital states.
In a more general approach, this approximation is not valid since contributions from bonding with neighboring atoms, and exchange interaction with core $s$-orbitals of the TM will lead to a non-zero electron density at the nucleus. Exchange interactions with the $s$ orbitals is symmetry forbidden for a $\Gamma_{5,6}$ doublet, but may arise for a doublet transforming as $\Gamma_4$.
This approximation does not, however, change the main conclusions of this work.
When presenting our results, we will address the consequences of this approximation in more detail.

In our numerical calculations, we always consider a nuclear spin-$5/2$.
We do this for simplicity and lean presentation in figures, since it reduces the number of electron-nucleus coupled states to consider, while it does not qualitatively change the results obtained.
Note that the relevant isotopes for Mo and V defects have nuclear spins $5/2$ and $7/2$, respectively.

Finally, we note that the electronic spin can also interact with nuclear spins of neighboring lattice Si or C atoms. However, within an ensemble of TM impurities, only a fraction of the defects have a neighboring non-zero nuclear spin. Additionally, defects with different nuclear spin environments are spectrally resolvable \cite{Wolfowicz2020}, and SiC can be fabricated with high isotopic purity, such that one can in principle choose to experimentally address defects with only spin-zero neighboring nuclear spins.

\subsection{Zeeman interaction}

The Zeeman Hamiltonian acts on the spin-orbit coupled states via a Hamiltonian of the type

\begin{equation}
H_{\text{Zee}} = -\frac{\mu_B}{h} \mathbf{B} \cdot (k\mathbf{L} + g_e\mathbf{S}) - \frac{g_n \mu_N}{h} \mathbf{B} \cdot \mathbf{I}
\end{equation}

\noindent  where $\mathbf{B}$ is the applied magnetic field with components $B_x$, $B_y$ and $B_z$, and magnitude $B$.

At small magnetic fields (such that $ B \mu_B/h  \ll \lambda k$) this Hamiltonian takes the shape of an effective-spin Hamiltonian for each Kramers doublet of the kind

\begin{equation}
H_{\text{Zee}} = -\frac{\mu_B}{h} \big(g_\parallel B_z \cdot \tilde{S}_z + g_\bot (B_x \cdot \tilde{S}_x + B_y \cdot \tilde{S}_y) \big) - \frac{g_n \mu_N}{h} \mathbf{B} \cdot \mathbf{I}
\end{equation}

\noindent where $\tilde{\mathbf{S}}$ is the effective spin-$1/2$ operator. 

\subsection{Numerical and analytical methods}

Below, we present both numerical and analytical results based on the model presented so far.
On the one hand, we numerically obtain the matrix elements for various time-dependent perturbations arising from the interaction between the defects' electric and magnetic dipole moments and oscillating electric and magnetic fields.
We achieve this by calculating these perturbations in terms of the basis states defined by the static Hamiltonian of equation~(\ref{eq::H_all}).
On the other hand, we obtain the analytical effective-spin Hamiltonians associated with each KD by investigating the effect of the hyperfine and Zeeman interactions on each KD independently.
The basis states for each KD are obtained by diagonalizing the crystal-field Hamiltonian (see \cite[section~I]{SI}).
We include SOC to zeroth-order, that is, we consider how it breaks the degeneracy between the two ground-state KDs, but disregard how it mixes ground and optically-excited manifolds.

\section{Results}

\subsection{Parameters consistent with experiment}

In the model presented so far, based on equation~(\ref{eq::H_all}), there are several free-parameters ($\Delta$, $\eta$, $\Delta_{\text{A}_1}$, $k$ and $\lambda$) which must be obtained either from first-principles calculations or from comparison with experimental results.
We choose to take the latter approach to restrict the parameter space where our model effectively describes the ground states of Mo and V defects.
We focus on V defects at the $\alpha$ site in $4$H-SiC \cite{kunzer1993,spindlberger2019}, since these are the defect centers most similar to Mo defects, which have promising features for quantum applications such as long spin-relaxation times in the ground-state \cite{Gilardoni2020}.
Additionally, the experimental parameters for this particular defect are well established, with reproducible measurements from various groups being present in literature \cite{baur1997,Wolfowicz2020}.
A similar analysis for other V configurations can be found in the supplementary information, \cite[section~V]{SI}.

The parameters $\Delta$ and $\Delta_{\text{A}_1}$ are only relevant to determine the energy difference between ground and optically excited states, and the first-order correction to the electronic wavefunctions due to spin-orbit coupling.
For this reason, we fix $\Delta$ at 1~eV, the order of magnitude of the optical energy splitting. We choose $\Delta_{\text{A}_1}$ to be on the order of 10~meV, but this choice has no consequence for the results obtained in the following.
In contrast, the parameters $\eta$, $k$ and $\lambda$ have significant implications for the ground-state properties of these defect centers.
In order to determine these parameters, we compare the energy eigenstates obtained from diagonalization of the Hamiltonian in equation~(\ref{eq::H_all}) (without hyperfine coupling) with the experimental parameters reported in Ref. \cite{Wolfowicz2020} for the splitting between the two ground-state KDs ($\Delta_{GS}$) and the Zeeman splitting of the lowest KD (determined by $g_\parallel$ and $g_\bot$). We consider a static magnetic field $B_0 = 100$~mT.
We do this by finding regions where combinations $k$, $\eta$ and $\lambda$ give ground state energy splittings that coincide with experiments within the experimental accuracy.

This agreement exists for $k$ values between $0.18$ and $0.37$, and is shown graphically for $k=0.3$ in figure~\ref{Fig::Model}(c). An \textit{ab-initio} approach that takes into account hybridization with orbitals other than the $d$ orbitals as well as dynamic Jahn-Teller effects gives an orbital reduction factor between 0.3 and 0.4 \cite{Csore2020}.
We find that using other values for $k$ from this range does not alter our general conclusions, and only leads to small shifts in the relevant numerical estimates for the hyperfine coupling parameters (see \cite[section~II]{SI}).
Lower $k$ values require increasing $|\eta|$ to match experimental results, whereas the spin-orbit coupling parameter $\lambda$ remains nearly constant (see SI for analogous plots for other values of $k$).
At $k=0.3$, $\eta \approx -0.4$ and $\lambda \approx 15$ meV constitute the parameter range that reliably reproduces the experimental properties of the ground-state electronic spin of this defect.
From here on in the text, we assume these parameters, and use these to generate the results presented in figure~\ref{Fig::Matrix} and figure~\ref{Fig::Energy}.

We note that the negative sign of $\eta$ depends on the convention taken for the energy splitting $\Delta_E$. A negative $\eta$ implies that the trigonal crystal field causes the orbitals $d_{\pm1}$ to have the lowest energy (see also figure~\ref{Fig::Model}(b)), possibly due to bonding or to the spatial distribution of electrons from the neighboring atoms \cite{spindlberger2019,csore2016}. 
This configuration leads to a ground state Kramers doublet that transforms as the $\Gamma_4$ irreducible representation of the double group $\bar{C}_{3v}$. 
The analytical effective-spin Hamiltonians obtained in parallel by us in Sec.~\ref{Sec::analytical} and by Tissot \textit{et al.} in \cite{Tissot2021Hyperf} for a doublet transforming as $\Gamma_4$ are in agreement with the experimental spectra of these V centers \cite{Wolfowicz2020}.

Figure~\ref{Fig::Model}(d) presents for these values of $k$, $\lambda$ and $\eta$ the approximate electronic wavefunction of  the ground state KD (and shows reasonable agreement with the wavefunction from reference~\cite{Csore2020}). 
In this graphical representation of the wavefunction in cartesian space, the azimuthal phase of the wavefunction at a certain point in space corresponds to its color according to the scale bar in figure~\ref{Fig::Matrix}(c).
The wavefunction clearly has a winding azimuthal phase, corresponding to a non-zero orbital angular momentum parallel to the three-fold rotational axis.
In this configuration, both ground state KDs have $g_\bot$ that would be measured as effectively $0$ at low static magnetic fields (below $1$ T).
In contrast, a magnetic field parallel to the crystal c-axis breaks the degeneracy of the KDs, giving rise to eigenstates where the effective electronic spin points parallel or antiparallel to the magnetic field.
From here on, we denote these eigenstates as $\ket{\Gamma_i \uparrow \downarrow}$, where $\Gamma_i$ corresponds to the symmetry of the KD, and the arrows indicate the effective spin direction.

{\begin{figure*}[t]
		\includegraphics[width=\textwidth]{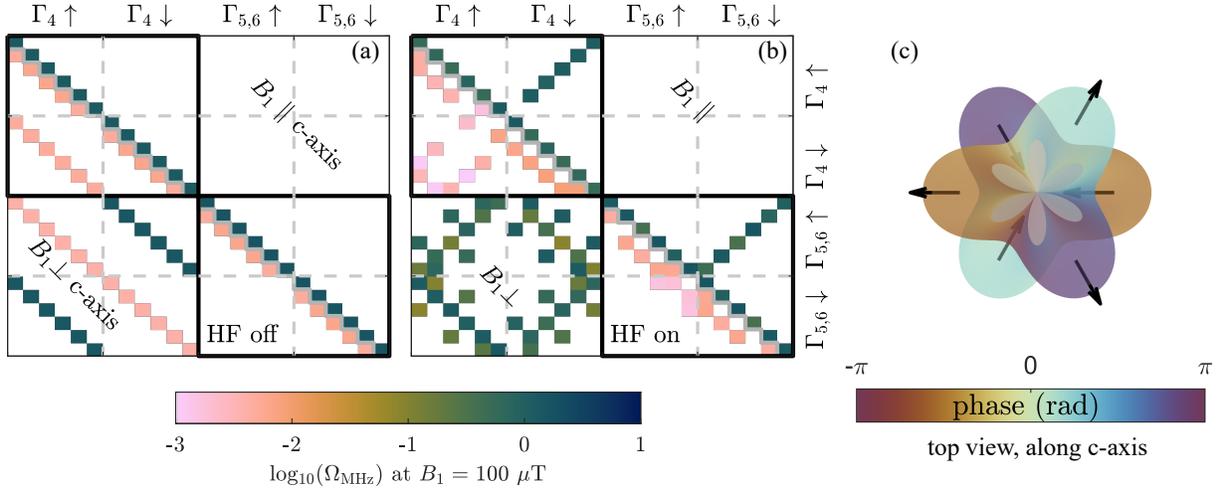}
		\caption{(a) Matrix elements due to an oscillating magnetic field parallel (diagonal and upper triangular elements) or perpendicular (lower triangular elements) to the c-axis, in the basis defined by the crystal field, spin-orbit coupling and a small magnetic field ($20$mT) along the c-axis, for the two ground state Kramers doublets ($\ket{\Gamma_4,\uparrow\downarrow}$ and $\ket{\Gamma_{5,6},\uparrow\downarrow}$), without hyperfine coupling. White matrix elements correspond to zero. Within each KD (contoured block diagonal parts of the full matrix), a magnetic field parallel to the c-axis does not lead to magnetic microwave transitions, consistent with a picture where the electronic spin is fully quantized along the static magnetic field direction. A magnetic field perpendicular to the c-axis leads to very weak electron and nuclear spin transitions in the $\Gamma_4$ doublet, and only nuclear spin transitions in the $\Gamma_{5,6}$ doublet. (b) Same as (a), but with the hyperfine interaction between electron and the central nuclear spin added to the static Hamiltonian. In this case, the electronic spin can flip via magnetic microwave transitions driven by an oscillating field parallel to the c-axis. (c) Top view of the wavefunction presented in figure~\ref{Fig::Model}(d). The arrows indicate the direction of the dipolar field caused by the central nuclear spin at the points of high electron density. For this particular electron distribution, the dipolar part of the hyperfine interaction favors a perpendicular orientation of the spins, such that the eigenstates are superpositions of up and down effective-electronic spin.
			\label{Fig::Matrix}}
	\end{figure*}
}

\subsection{Interaction with magnetic component of microwave fields}

In order to determine which spin transitions can be driven by oscillating magnetic fields, we numerically investigate the matrix elements of the Zeeman operator between the effective-spin eigenstates associated with a Hamiltonian including the crystal-field potential, spin-orbit coupling and the interaction with a static magnetic field ($B_0 = 20$~mT) parallel to the c-axis (figure~\ref{Fig::Matrix}(a,b)).
We focus on the two ground-state KDs, such that the basis used in the matrix representation corresponds to the $24$ lowest eigenstates associated with the two of effective-spin-$1/2$ doublets, coupled to a nucleus with spin-$5/2$.
We investigate the matrix elements arising from the interaction with an oscillating magnetic field parallel (diagonal elements, and upper-triangular elements) and perpendicular (lower-triangular elements) to the static magnetic field and the symmetry axis of the crystal.
Additionally, in figure~\ref{Fig::Matrix}(b) we include hyperfine interaction to the static Hamiltonian.

When hyperfine coupling between the electron and nuclear spin is not present, the $\Gamma_{5,6}$ doublet (bottom right block diagonal) is fully protected from effective-spin transitions due to oscillating magnetic fields (figure~\ref{Fig::Matrix}(a)).
This is characterized by the absence of off-diagonal elements between states $\ket{\Gamma_{5,6} \uparrow}$ and $\ket{\Gamma_{5,6} \downarrow}$ when considering magnetic fields in either direction.
Magnetic fields perpendicular to the defect symmetry axis may lead to nuclear-resonance transitions, as well as transitions between the $\Gamma_{5,6}$ and $\Gamma_{4}$ doublets.
However, driving these latter transitions requires oscillating fiels in the sub-THz range.
In contrast, within the $\Gamma_4$ doublet, a magnetic field perpendicular to the symmetry axis of the defect center gives rise to matrix elements between the $\ket{\Gamma_{4} \uparrow}$ and $\ket{\Gamma_{4} \downarrow}$ due to the small but non-vanishing $g_\bot$.
These resonances, however, have very small Rabi frequencies due to the very weak interaction with magnetic fields in the plane.
For an oscillating magnetic field with magnitude $B_1 = 100~\mu$T, we expect a Rabi frequency on the order of $1$ kHz.

When we include the hyperfine interaction with the central nucleus in the static Hamiltonian that determines the basis states for representation of the Zeeman perturbations, the matrix representation of the Zeeman operators shows significant changes (figure~\ref{Fig::Matrix}(b)). Most notably, off-diagonal matrix elements appear for interaction with oscillating magnetic fields parallel to the symmetry axis of the defect (and the static magnetic field), for both ground-state effective-spin doublets.
This indicates that the hyperfine coupling leads to mixing between the two effective-electronic-spin eigenstates $\ket{\Gamma_i \uparrow}$ and $\ket{\Gamma_i \downarrow}$.
An oscillating magnetic field parallel to the symmetry axis of the defect modulates the amount of mixing, and this leads to efficient effective-spin transitions.
This hyperfine-induced mixing arises from the anisotropic hyperfine interaction.

In our model, the main contribution to this anisotropic interaction is the dipolar coupling between nuclear and electronic magnetic-dipole moments.
Figures~\ref{Fig::Model}(d) and \ref{Fig::Matrix}(c) show the spatial configuration of the ground-state electronic wavefunction.
We indicate with arrows the direction of the dipolar field at the points of high electron density due to a magnetic moment at the origin and parallel to the symmetry axis of the defect center.
Due to the particular spatial distribution of the electronic wavefunction around the nucleus, this dipolar field tends to align the electronic and nuclear magnetic moments perpendicular to each other.
A magnetic field along the symmetry axis of the defect, however, favors the alignment of both magnetic moments along its direction.
These two effects compete, leading to a strong dependency of the eigenstate composition on the magnitude of the static mangetic field when the hyperfine and electronic-Zeeman energies are comparable (figure~\ref{Fig::Energy}(a)).
Thus, the influence of hyperfine is strongly dependent on the magnitude of the static magnetic field, and is most relevant for magnetic fields below 100~mT.
Figure~\ref{Fig::Energy} presents this for the $\Gamma_4$ KD, which is lowest in energy.
As a consequence, the Rabi frequencies for resonant driving with oscillating magnetic fields along the crystal c-axis decrease significantly as the magnitude of the static field increases.
At low static magnetic fields (20~mT), Rabi frequencies as high as 1~MHz can be obtained for oscillating fields of $B_1 = 100~\mu$T.
At a static magnetic field of 100~mT, this value decreases by an order of magnitude (figure~\ref{Fig::Energy}(b)).
In contrast, when the oscillating field is perpendicular to the crystal c-axis and the static magnetic field, effective spin-transitions within the $\Gamma_4$ doublet arise from the non-zero $g_\bot$, and are very small but largely independent from the magnitude of the static magnetic field~(figure~\ref{Fig::Energy}(c)).

{\begin{figure*}[t]
		\includegraphics[width=\textwidth]{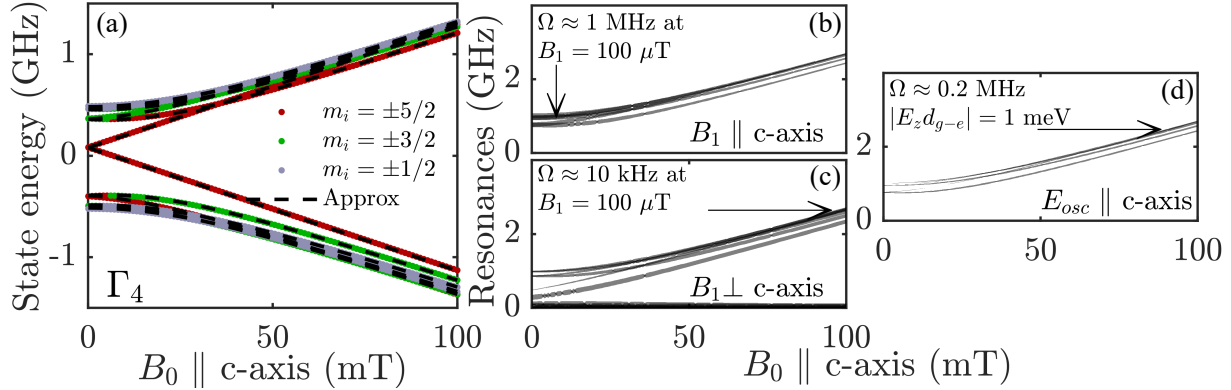}
		\caption{(a) Energies of the $12$ eigenstates corresponding to the $\Gamma_4$ ground-state electronic spin doublet, hyperfine-coupled to a central nucleus with a spin-$5/2$, as a function of the static magnetic field $B_0$ applied parallel to the c-axis. Results were obtained numerically (colored lines), and by diagonalizing the effective spin Hamiltonian associated with a $\Gamma_4$ doublet (dashed line). (b-d) Microwave resonance lines obtained numerically for an oscillating magnetic field parallel (b) or perpendicular (c) to the c-axis, or an oscillating electric field parallel (d) to the c-axis. The thickness of the resonance lines correspond to their Rabi frequencies, with different scales in each figure. Scales are provided by the labels and arrows, indicating the Rabi frequency at a certain point. (b) If the static field is small (Zeeman splitting comparable to hyperfine coupling strength), a small oscillating magnetic field ($B_1 \approx 100~\mu$T) parallel to the c-axis leads to very efficient electronic spin transitions with Rabi frequencies on the order of $1$~MHz. These resonances weaken significantly as the static magnetic field increases. (c) In a $\Gamma_4$ doublet, direct interaction with an oscillating field perpendicular to the c-axis leads to electronic microwave transitions with Rabi frequencies that are largely independent of the applied static field, but small due to the reduced $g_\bot$. (d) A large enough electric field parallel to the c-axis mixes ground and optically excited states and deforms the electron density, modulating the influence of the anisotropic dipolar hyperfine coupling and leading to electronic spin transitions.
			\label{Fig::Energy}}
	\end{figure*}
}

\subsection{Interaction with electric component of microwave fields}

We now turn to considering whether the electric component of a resonant microwave field can also drive spin transitions within the ground-state KD.
One way to consider the interaction of the ground-state KDs with oscillating electric fields is to investigate the matrix representation of the Hamiltonian

\begin{equation}
	H_\text{el} = -e \vec{E} \cdot \vec{r}
\end{equation}

\noindent where $e$ is the electron charge and $\vec{E}$ is the electric-field vector, and $H_\text{el}$ is obtained in the dipolar approximation.
However, when we only consider the $d$-orbitals of the TM, this operator has all zero matrix elements due to the spatial parity of these orbitals.
Nonetheless, we know that the defect center does not have inversion symmetry, such that dipolar interactions with electric fields are in principle symmetry-allowed and will likely arise due to mixing with other orbital states ($p$-orbitals, for example). Additionally, the defect shows bright optical transitions between ground and optically excited states, such that the matrix elements of the electric dipole moment between these states must be non-zero.
To remedy this discrepancy we include the interaction with electric fields parallel to the symmetry axis of the crystal by considering an extra contribution to the crystal field potential via a change of the parameter $\eta$.

Hence, we model an electric field in this direction as an additional contribution to the crystal field that alters the relative strengths between the original tetrahedral (figure~\ref{Fig::Model}(a)) and trigonal fields (figura~\ref{Fig::Model}(b)).
Although an interaction of this type does not change the symmetry of the eigenstates, it does lead to mixing between ground and optically excited states.
In turn, this changes the spatial distribution of the electron density.
Figures~\ref{Fig::Model}(d) and \ref{Fig::Matrix}(c) show that the hyperfine interaction depends strongly on the spatial configuration of the electronic wavefunctions.
Thus, an electric field along this direction modulates the relative strength of the isotropic and anisotropic components of the hyperfine interaction via modulating the spatial configuration of the electronic wavefunction.
Overall, this causes that an electric field drives effective-spin transitions within the ground-state KDs (figure~\ref{Fig::Energy}(c)).
Notably, this effect only arises if hyperfine coupling is present. In its absence, effective-spin transitions are protected by Kramers theorem, even upon modulations of $\eta$. 

Since these transitions arise from mixing between the ground and optically excited states, the Rabi frequencies associated with this effect depend linearly on the ratio between the additional $C_{3v}$ contribution to the crystal field and the original splitting between ground and optically excited state (approximately $\Delta$).
Numerically, we estimate that a trigonal contribution of approximately 1~meV will give rise to Rabi frequencies of hundreds of kHz (figure~\ref{Fig::Energy}(d) and \cite[section~III]{SI}).
In order to estimate the magnitude of the electric fields necessary to drive the spin transitions at these rates, we need to have a reliable estimate for the relevant transition dipole moment $d_{g-e} = -e \braket{\Gamma_i^g | z}{\Gamma_i^e}$, where $\Gamma_i^{g,e}$ correspond to the $\Gamma_i$ eigenstates in ground and optically excited states, respectively.
This can be determined experimentally or via first-principles calculations.
Given that the TM defects have radiative lifetimes on order 100~ns \cite{spindlberger2019,Wolfowicz2020}, a conservative estimate for the transition dipole moment gives a magnitude of about $d_{g-e} \approx 1~\text{D} \approx 0.2~e \text{\AA}$.
For this value, an oscillating electric field with magnitude $E_z = 50$~V/$\mu$m will drive effective-spin transitions at these rates.

\subsection{Analytical expressions and effective-spin Hamiltonians}
\label{Sec::analytical}

We can obtain analytical effective-spin Hamiltonians for the two ground state KDs based on the model presented so far.
These expressions can be fit to experiments, directly connecting the theory developed here to experimental observations.
In order to do this, we diagonalize the crystal field Hamiltonian including both the tetrahedral and trigonal contributions, and focus on the two lowest Kramers doublets. 
(This is done explicitly in \cite[section~VI]{SI}.)
We include spin-orbit coupling in zeroth order -- that is, we consider the energy splitting between the two KDs, but disregard the mixing between ground and optically excited states caused by SOC.
Finally, we investigate the effect of the hyperfine and Zeeman operators within each KD.
We obtain effective-spin Hamiltonians of the type

\begin{align}
H_{\Gamma_4} &= H_{\text{HF}}^{eff} +  H_{\text{Zee,el}}^{eff} +  H_{\text{Zee,nuc}}^{eff} \\
&H_{\text{HF}}^{eff} = a_{\parallel,\Gamma_4} \tilde{S}_z I_z + a_{\bot,\Gamma_4} (\tilde{S}_+ I_+ + \tilde{S}_- I_-)\\
&H_{\text{Zee,el}}^{eff} = -\mu_B \big(g_{\parallel,\Gamma_4} B_z \tilde{S}_z + g_{\bot,\Gamma_4} (B_x \tilde{S}_x + B_y \tilde{S}_y)\big)\\
&H_{\text{Zee,nuc}}^{eff} = -\mu_N g_n \big(B_z I_z + B_x I_x + B_y I_y\big)
\end{align}

\begin{align}
H_{\Gamma_{5,6}} &= H_{\text{HF}}^{eff} +  H_{\text{Zee,el}}^{eff} +  H_{\text{Zee,nuc}}^{eff} \\
&H_{\text{HF}}^{eff} = (a_{\parallel,\Gamma_{5,6}} \tilde{S}_z + a_{\bot,\Gamma_{5,6}} \tilde{S}_y) I_z\\
&H_{\text{Zee,el}}^{eff} = -\mu_B g_{\parallel,\Gamma_{5,6}} B_z \tilde{S}_z\\
&H_{\text{Zee,nuc}}^{eff} = -\mu_N g_n \big(B_z I_z + B_x I_x + B_y I_y\big)
\end{align}

The approximate dependency of the parameters $a_{\parallel,\bot}$, $g_{\parallel,\bot}$ on the parameters describing the microscopic configuration of the system, $k$, $\eta$ and $\Delta$, is given in the supplementary information \cite[section~VI]{SI}.
The approximate energy levels obtained from these effective-spin Hamiltonians for the lowest-energy KD are plotted in figure~\ref{Fig::Energy}(a), dashed lines, and overlap well with the numerical results.

These effective-spin Hamiltonians are clearly different for KDs with different symmetries, and give rise to spectra that are clearly distinguishable (see \cite[section~III]{SI} for the energies and magnetic resonance spectra of a KD transforming as $\Gamma_{5,6}$). Notably, microwave resonance spectra of $\Gamma_4$ doublets due to interaction with oscillating magnetic fields perpendicular to the crystal c-axis shows evidence of a pair of hyperfine sublevels whose energy depends linearly on the applied static magnetic field (figures~\ref{Fig::Energy}(a,c)). This is absent for $\Gamma_{5,6}$ doublets (SI). Thus, hyperfine resolved magnetic resonance spectra of these defects' ground-state KDs can be used to determine their symmetry properties.
As mentioned before, comparison with experiments shows that the lowest-energy KD of V defects in SiC transforms as $\Gamma_4$.
For the configuration considered here, we find for this doublet $a_\parallel \approx -63$ MHz, and $a_\bot \approx 332$ MHz.
When compared to the experimentally measured values, our model overestimates $|a_\bot|$ and underestimates $|a_\parallel|$ \cite{Tissot2021Hyperf}.
This could be explained by the fact that in our model we overlook the hyperfine Fermi-contact interaction and contributions from $s$-orbitals to the wavefunctions, which is in principle allowed for states transforming as $\Gamma_4$. The latter gives rise to an isotropic contribution with $a_\parallel \approx a_\bot$.
This could be confirmed by first-principle calculations, for example.

\section{Discussion and conclusion}

The results of the previous section show that the apparent discrepancy between the symmetry protection of the TM spin-$1/2$ states (with respect to electronic spin transitions) and the observed magnetic resonance results on V defects can be fully resolved by considering the interaction with a central nuclear spin.
Our model, focused on an electron localized in the $d$-orbitals of the TM core, encompasses this behavior and shines light on the underlying physical mechanisms.
For defects with a central nuclear spin, the hyperfine Hamiltonian is a combination of an isotropic term -- that favors parallel alignment of the nuclear spin and electronic effective spin -- and an anisotropic term -- that favors perpendicular alignment of these magnetic moments.
The latter leads to mixing between electronic spin sublevels.
Periodic modulation of the degree of this mixing by microwaves enables transitions, when the oscillating electric and magnetic fields are parallel to the symmetry axis of the defect.
This anisotropic hyperfine interaction arises mostly from the classical dipolar coupling between the magnetic moments of the nucleus and electron, which is strongly dependent on the spatial configuration of the wavefunction.
This dependency also means that the strength of the hyperfine coupling -- and consequently also the electronic spin eigenstates -- is sensitive to local electric and magnetic inhomogeneities.
This additional hyperfine induced mixing of electronic spins may increase the rate of direct spin flips due to interactions with lattice phonons or with neighboring paramagnetic impurities, for example, possibly limiting the spin relaxation at low magnetic fields.

This conclusion is still consistent, however, with the seconds-long spin relaxation times measured for an ensemble of Mo defects \cite{Gilardoni2020} at temperatures where direct spin-flips dominate the spin-relaxation processes.
In this case, the most abundant Mo isotopes ($\sim$75$\%$) have zero nuclear spin, such that hyperfine coupling does not play a role for the majority of the ensemble.
Additionally, direct spin flips are not the limiting process leading to spin relaxation at low temperatures for these defects, such that these hyperfine-mediated spin flips are still likely to be slower than the seconds-long spin relaxation times.
For V ensembles, where all of the naturally occurring isotopes have a non-zero nuclear spin, hyperfine coupling to the central nucleus is always relevant.
On the one hand, this coupling enables efficient microwave control of the electronic spin, which can subsequently be read out optically.
On the other hand, these hyperfine-enabled electronic spin transitions due to oscillating magnetic and electric fields may enhance the defect sensitivity to neighboring spins and lattice phonons, respectively. This may play a role on determining the spin relaxation times of these defect centers at low magnetic fields \cite{abragam1970}, at temperatures where direct spin-flips dominate the spin relaxation processes.
For this reason, a full characterization of the spin relaxation time of V defects at low magnetic fields and temperatures remains relevant to determine whether there is indeed a trade-off between efficient spin control and slow spin relaxation.
This will allow to identify optimal ranges or protocols for operation of these defect centers as optically active spin-qubits, where the defect centers are still addressable but robust against environment-mediated spin-relaxation.

For quantum operation, not only the spin-relaxation times, but also spin-coherence times have significant implications.
Our results indicate that the presence of hyperfine coupling may be responsible for extending the latter.
The competing behavior of the hyperfine and the Zeeman interactions gives rise to energy eigenstates with anticrossing points.
For the particular defect configuration and KD depicted in figure~\ref{Fig::Energy}(a), these anticrossing points happen at zero magnetic field, and thus provide clock transitions at these points \cite{Wolfowicz2013,Gimenez2020,Shiddiq2016}.
There, the energy levels are not dependent (to first order) on small variations of the magnetic field.
This leads to a suppression of decoherence and dephasing processes related to spectral diffusion, which results in prolonged coherence times for the quantum states.
For defects in SiC, these processes are already partly reduced at finite magnetic fields (when compared to silicon or diamond, for example) since the lattice contains two different types of nuclear spins, and flip-flop processes between lattice nuclei are rare \cite{Seo2016,Yang2014}.
Nonetheless, the small dependence of the energy eigenlevels on the magnetic field magnitude means that, in this configuration, the defects are even less sensitive to the existing flip-flop events between lattice nuclei.
In samples with high defect concentration, where individual defect centers feel each other, however, this decoherence protection is expected to be most relevant, leading to increased coherence close to these anticrossing configurations.

The coupling between the optically addressable electron spin and the potentially long-lived nuclear spin provides a key ingredient for advanced quantum communication and computation architectures \cite{Nemoto2014,Nguyen2019,Awschalom2021}. In the ground-state eigenstates, electronic and nuclear spins are fully entangled \cite{Tissot2021Hyperf}. 
At low magnetic fields and for the $\Gamma_4$ doublets, the eigenstates correspond to superpositions of nuclear spin states, which may impact the nuclear spin coherence lifetime. 
The detailed understanding of the coupling to external fields developed herein calls for experimental studies of the coherence properties, and the development of bespoke decoupling methods to protect the quantum states.
Conversely, these characteristics offer a wealth of opportunities for the development of control methods which leverage the large Hilbert space within the electron-nuclear spin manifold:
the possibility of driving spin transitions with oscillating electric fields along the symmetry axis of the defect centers is technologically relevant, and may enable very efficient device architectures \cite{Klimov2014,falk2014,Anderson2019,Widmann2019}.
Oscillating electric fields along this direction can be applied with high amplitudes and in a very homogeneous manner in devices with parallel plate capacitors, or in monolithic p-i-n junctions with defects embedded in the intrinsic layer.
Additionally, these results also indicate that strain modulation can be used for electronic-spin control, enabling hybrid architectures where these defect centers are entangled with mechanical oscillation modes \cite{Whiteley2019}.

\section*{Acknloweledgements}
We thank J. Hendriks, T. Bosma, and A. Gali for discussions. We thank B. Tissot and G. Burkard for providing a preliminary version of their related manuscript. Financial support was provided by the Zernike Institute BIS program, and the EU H2020 project QuanTELCO (862721).

\section*{Author contributions}
The project was initiated by CMG and CHvdW. II and FH contributed with theory and simulation work. All authors contributed via discussions, and read and commented on the manuscript.
\bibliography{myBib}

\end{document}